\documentclass[12pt, notitlepage]{article}

\usepackage{a4}

\usepackage{xcolor}

\definecolor{TUMblue}{RGB}{0, 101, 189}
\definecolor{TUMlightblue}{RGB}{100,160,200}
\definecolor{TUMgreen}{RGB}{162,173,0}
\definecolor{TUMorange}{RGB}{227,114,034}
\definecolor{TUMivory}{RGB}{218,215,203}

\usepackage{footnote}

\usepackage{natbib}

\usepackage{hyperref}
\hypersetup{
	colorlinks=true,
	linkcolor=TUMorange,
	citecolor=TUMorange,
	filecolor=TUMorange,
	urlcolor=TUMorange
}

\usepackage{etoolbox}
\makeatletter

\pretocmd{\NAT@citex}{%
	\let\NAT@hyper@\NAT@hyper@citex
	\def\NAT@postnote{#2}%
	\setcounter{NAT@total@cites}{0}%
	\setcounter{NAT@count@cites}{0}%
	\forcsvlist{\stepcounter{NAT@total@cites}\@gobble}{#3}}{}{}
\newcounter{NAT@total@cites}
\newcounter{NAT@count@cites}
\def\NAT@postnote{}

\def\NAT@hyper@citex#1{%
	\stepcounter{NAT@count@cites}%
	\hyper@natlinkstart{\@citeb\@extra@b@citeb}#1%
	\ifnumequal{\value{NAT@count@cites}}{\value{NAT@total@cites}}
	{\ifNAT@swa\else\if*\NAT@postnote*\else%
		\NAT@cmt\NAT@postnote\global\def\NAT@postnote{}\fi\fi}{}%
	\ifNAT@swa\else\if\relax\NAT@date\relax
	\else\NAT@@close\global\let\NAT@nm\@empty\fi\fi
	\hyper@natlinkend}
\renewcommand\hyper@natlinkbreak[2]{#1}

\patchcmd{\NAT@citex}
{\ifNAT@swa\else\if*#2*\else\NAT@cmt#2\fi
	\if\relax\NAT@date\relax\else\NAT@@close\fi\fi}{}{}{}
\patchcmd{\NAT@citex}
{\if\relax\NAT@date\relax\NAT@def@citea\else\NAT@def@citea@close\fi}
{\if\relax\NAT@date\relax\NAT@def@citea\else\NAT@def@citea@space\fi}{}{}

\makeatother

\usepackage{amssymb, graphicx}
\usepackage{subfigure}
\usepackage{amsmath}
\usepackage{amsthm}
\usepackage{undertilde}
\usepackage{verbatim}
\usepackage{bbm}
\usepackage{ dsfont }
\usepackage{geometry}
\usepackage{pdflscape}
\usepackage{multirow}
\usepackage{aliascnt}

\newcommand{\mynewtheorem}[2]{
	\newaliascnt{#1}{dummy}
	\newtheorem{#1}[#1]{#2}
	\aliascntresetthe{#1}
	\expandafter\def\csname #1autorefname\endcsname{#2}
}

\theoremstyle{definition}
\mynewtheorem{thm}{Theorem}
\mynewtheorem{defi}{Definition}
\mynewtheorem{lem}{Lemma}
\mynewtheorem{cor}{Corollary}
\mynewtheorem{prop}{Proposition}
\mynewtheorem{exa}{Example}
\mynewtheorem{alg}{Algorithm}
\mynewtheorem{rem}{Remark}
\mynewtheorem{bsp}{Example}

\def\equationautorefname~#1\null{Equation~(#1)\null}

\usepackage{sectsty}
\allsectionsfont{\sffamily}

\usepackage{url}

\usepackage{caption}
\newcommand{\Rbb}{\mathbb{R}}

\newcommand*\diff{\mathop{}\!\mathrm{d}}

\newcommand{\thetab}{\boldsymbol{\theta}}

\newcommand{\Ub}{\mathbf{U}}

\newcommand{\Zb}{\mathbf{Z}}
\newcommand{\zb}{\mathbf{z}}

\newcommand{\Cc}{\mathcal{C}}

\newcommand{\Fc}{\mathcal{F}}
\newcommand{\Gc}{\mathcal{G}}

\newcommand{\Jc}{\mathcal{J}}

\newcommand{\Nc}{\mathcal{N}}

\newcommand{\Tc}{\mathcal{T}}

\newcommand{\be}{\begin{equation}}
	\newcommand{\ee}{\end{equation}}

\DeclareMathOperator{\BBa}{BB1}
\DeclareMathOperator{\BBb}{BB6}

\DeclareMathOperator{\BBd}{BB8}

\DeclareMathOperator{\sgn}{sgn}
\DeclareMathOperator{\tcop}{t}

\begin{document}

{
	\renewcommand*{\thefootnote}{\fnsymbol{footnote}}
	\title{\textbf{\sffamily Examination and visualisation of the simplifying assumption for vine copulas in three dimensions
		}}
		\date{\small \today}
		\author{Matthias Killiches\footnote{Zentrum Mathematik, Technische Universit\"at M\"unchen, Boltzmannstra\ss e 3, 85748 Garching, Germany.}\and Daniel Kraus$^*$\footnote{Corresponding author: \texttt{daniel.kraus@tum.de}.} \and Claudia Czado$^*$}	
		\maketitle
		
		\begin{abstract}
			Vine copulas are a highly flexible class of dependence models, which are based on the decomposition of the density into bivariate building blocks. For applications one usually makes the simplifying assumption that copulas of conditional distributions are independent of the variables on which they are conditioned. However this assumption has been criticised for being too restrictive. We examine both simplified and non-simplified vine copulas in three dimensions and investigate conceptual differences. We show and compare contour surfaces of three-dimensional vine copula models, which prove to be much more informative than the contour lines of the bivariate marginals. Our investigation shows that non-simplified vine copulas can exhibit arbitrarily irregular shapes, whereas simplified vine copulas appear to be smooth extrapolations of their bivariate margins to three dimensions. In addition to a variety of constructed examples, we also investigate a three-dimensional subset of the well-known uranium data set and visually detect that a non-simplified vine copula is necessary to capture its complex dependence structure.\\
			
		\end{abstract}

\section{Introduction}\label{sec:Intro}
Dependence modeling has become a growing research area of high interest in the last decades. In finance, regulatory requirements like Basel III \citep{basel2009global} and Solvency II \citep{europe2009solvency} have increased the need for sophisticated risk assessment, thus making a proper understanding of the interdependencies between different quantities inevitable. For hydrological applications the dependence between rainfall, wind speed and other parameters is crucial for setting up appropriate models. Basically, dependence plays a role whenever there is more than one source of randomness.\\
Among dependence models, copulas have taken a prominent role since they allow for separate modelling of marginal distributions and dependence structure \citep{Sklar}. A popular choice for modelling copulas are so-called \textit{vine copulas}, also known as \textit{pair-copula constructions}, which are based on a decomposition of the joint copula density into bivariate building blocks. Since the publication of the seminal papers by \cite{bedford2002vines} and \cite{aasczado}, these copulas have gained more and more popularity due to their flexibility and numerical applicability. When working with vine copulas one often makes the \textit{simplifying assumption} that pair-copulas of conditional distributions are independent of the values of the variables on which they are conditioned. Although enabling estimation and inference even in high dimensions, this assumption has also been criticised \citep[e.g.][]{acar2012beyond, spanhel2015simplified}. Our goal is to shed some light on the implications of this simplifying assumption by visualising the densities of simplified and non-simplified models. For this purpose, we concentrate on the three-dimensional case. This has the advantage that the corresponding pair-copula construction contains only one copula describing the dependence between conditional variables, making the interpretation of the results easier. Further it is possible to visualise three-dimensional densities by plotting their contour surfaces. We will see that these plots contain much more information than the bivariate contour lines of the three two-dimensional margins.\\
The paper is organised as follows: In \autoref{sec:VC} we provide an introduction to vine copulas and a formal definition of the simplifying assumption. We show visualisations of simplified vine copulas in \autoref{sec:Vis_Simp}, while \autoref{sec:Vis_Non_Simp} contains three-dimensional plots of non-simplified vine copulas and their simplified vine copula approximations. We present applications to simulated and real data in \autoref{sec:App}. The paper concludes with \autoref{sec:Conclusion}. In addition to the paper, we provide a web-application \cite[implemented using the \texttt{shiny} package by][]{shiny} for visualising three-dimensional vine copulas (\texttt{\url{https://vinecopula.shinyapps.io/Vine3DPlot}}).

\section{Vine copulas and the simplifying assumption}\label{sec:VC}
Since the seminal work of Sklar has been published \citep{Sklar} the concept of \textit{copulas} has become more and more popular in statistical modelling. Copulas are $d$-dimensional distribution functions on the hypercube $[0,1]^d$ with uniformly distributed margins. The following relationship proven in \textit{Sklar's Theorem} makes copulas extremely useful for statistical applications: Let $F\colon \Rbb^d \to [0,1]$ be the joint distribution function of a $d$-dimensional random variable $(X_1,\ldots,X_d)^\top$ with univariate marginal distribution functions $F_j$, $j=1,\ldots,d$. Then there exists a copula $C\colon [0,1]^d \to [0,1]$ such that
\begin{equation}\label{eq:sklar}
F(x_1,\ldots,x_d)=C\left(F_1(x_1),\ldots, F_d(x_d)\right).
\end{equation}
This copula $C$ is unique if all $X_j$ are continuous. If additionally the so-called \emph{copula density} 
\[
c(u_1,\ldots,u_d)=\frac{\partial^d}{\partial u_1 \cdots \partial u_d}C(u_1,\ldots,u_d)
\]
exists, we have 
\begin{equation}\label{eq:sklar_dens}
f(x_1,\ldots,x_d)=c\left(F_1(x_1),\ldots, F_d(x_d)\right)f_1(x_1)\cdots f_d(x_d),
\end{equation}
where $f_j$ are the marginal densities.
Conversely, one can define a multivariate distribution by specifying a $d$-dimensional copula and $d$ univariate marginal distributions with the help of \eqref{eq:sklar}. This means that the marginals and the dependence structure can be modelled separately. \cite{nelsen2006introduction} and \cite{joe1997multivariate} provide extensive treatments of theoretical and practical aspects of copulas.\\

For modelling purposes many copula classes have been developed, e.g.\ elliptical, Archimedean and extreme-value copulas. Usually the copula families in these classes are determined by a small number of parameters such that they are rather inflexible in high dimensions. This lack of flexibility has been overcome by \textit{vine copulas}. The concept of constructing copula densities using a combination of only bivariate building blocks was introduced in \cite{bedford2002vines}. Based on this work, \cite{aasczado} developed statistical inference methods using these vine copulas or \textit{pair-copula constructions} (PCCs). This class of copulas has been a frequent subject of recent research. \cite{dissmann2013selecting} presented a sequential estimation method for vine copulas. Vine models were used for quantile regression in \cite{noh2015semiparametric}, \cite{de2016semiparametric} and \cite{kraus2015d} and the dependence of finite block maxima of vine copulas was investigated in \cite{killiches2015maxima}. To extend the approach to non-continuous models \cite{panagiotelis2012pair} as well as \cite{stober2015comorbidity} studied the application of vine copulas to discrete data. Additionally several nonparametric methods for the estimation of the pair-copulas in a vine model have been developed, e.g.\ kernel density based estimation \citep{nagler2015evading}, estimation using splines \citep{kauermann2014flexible,schellhase2016estimating} and the empirical copula \citep{haff2015nonparametric}. Further there is a multitude of real data applications, especially in the context of finance \citep[e.g.][]{brechmann2014flexible,maya2015latin,brechmann2013risk}. Extended models for describing geo-spatial dependence were introduced in \cite{erhardt2015r} and \cite{erhardt2015spatial}. Vine copulas are particularly interesting for researchers and practitioners from all fields due to the \textsf{R} package \texttt{VineCopula} \citep{VC}, where algorithms for estimation, simulation and model diagnostics are implemented.

Since we will only consider three-dimensional examples in this paper, we introduce the concept of vine copulas in $d=3$ dimensions. For a general introduction to vine copulas we refer to \cite{aasczado} and \cite{stoeber2012}. Using Sklar's Theorem \eqref{eq:sklar_dens} for the conditional density $c_{13| 2}$---as has already been done by \cite[p.\ 533]{patton2006modelling}---and the fact that the one-dimensional marginals of a copula are uniformly distributed, i.e.\ $c_j(u_j)=1$ for any $u_j\in[0,1]$, we can decompose the copula density $c$ of a random vector $\Ub=(U_1,U_2,U_3)^\top$ with uniformly distributed margins $U_j$ as follows:
\begin{equation} \label{eq:3dvinedensity}
\begin{split}
c(u_1,u_2,u_3)&=c_{13| 2}(u_1,u_3| u_2)\,c_2(u_2)\\
&= c_{13;2}\left(C_{1| 2}(u_1| u_2),C_{3| 2}(u_3|u_2);u_2\right)c_{1|2}(u_1| u_2)\,c_{2| 3}(u_2| u_3)\\
&=c_{13;2}\left(C_{1|2}(u_1|u_2),C_{3|2}(u_3|u_2);u_2\right)c_{12}(u_1,u_2)\,c_{23}(u_2,u_3),
\end{split}
\end{equation}
where  $c_{13| 2}(\,\cdot\,,\cdot\,|u_2)$ is the density of the conditional distribution of $(U_1,U_3)|U_2=u_2$ and $c_{13;2}(\,\cdot\,,\cdot\,;u_2)$ is the copula density associated with the conditional distribution of $(U_1,U_3)|U_2=u_2$. The distribution function of the conditional distribution of $U_j$ given $U_2=u_2$ is denoted by $C_{j|2}(\,\cdot\,| u_2)$, $j=1,3$. It can be obtained by partial differentiation \citep{joe1997multivariate}: $C_{1|2}(u_1|u_2)=\partial/\partial u_2\,C_{12}(u_1,u_2)$ and $C_{3|2}(u_3|u_2)=\partial/\partial u_2\,C_{23}(u_2,u_3)$. Hence we can describe the entire copula density $c$ by specifying only three bivariate copulas. 

It would also have been possible to choose $U_1$ or $U_3$ as conditioning variable. Then we would have ended up with
\[
c(u_1,u_2,u_3)=c_{23;1}\left(C_{2|1}(u_2|u_1),C_{3|1}(u_3|u_1);u_1\right)c_{12}(u_1,u_2)\,c_{13}(u_1,u_3)
\]
or
\[
c(u_1,u_2,u_3)=c_{12;3}\left(C_{1|3}(u_1|u_3),C_{2|3}(u_2|u_3);u_3\right)c_{13}(u_1,u_3)\,c_{23}(u_2,u_3),
\]
respectively. However for our purpose all three structures are equivalent since they can be obtained by relabelling Therefore, throughout the paper, we will always use the decomposition from \eqref{eq:3dvinedensity}. Here the bivariate marginals $c_{12}$ and $c_{23}$ are explicitly specified. The third bivariate marginal $c_{13}$ is implicitly defined and can be obtained via one-dimensional integration:
\begin{equation}\label{eq:c13}
c_{13}(u_1,u_3)=\int_{0}^{1}c(u_1,v,u_3)\diff v.
\end{equation}

In general the bivariate copula density function $c_{13;2}(\,\cdot\,,\cdot\,;u_2)$ depends on the value of $u_2$. Yet when it comes to modelling, one often makes the so-called \textit{simplifying assumption} that $c_{13;2}$ does not depend on $u_2$, i.e.\
\[
c_{13;2}(u_1,u_3;u_2)=c_{13;2}(u_1,u_3)
\]
for all $u_1,u_2,u_3\in [0,1]$. To enable fast and robust inference this assumption is made in a multitude of cases. Nevertheless there has also been research on \textit{non-simplified vine copulas} and the question when the simplifying assumption is justified. \cite{stoeber2013simplified} investigated which multivariate copula models can be decomposed into a simplified vine copula. While \cite{haff2010simplified} came to the conclusion that simplified vine copulas are ``a rather good solution, even when the simplifying assumption is far from being fulfilled by the actual model'', \cite{acar2012beyond} criticised this statement as being too optimistic and show cases where non-simplified vine copulas are needed. In particular, they looked at a three-dimensional subset of the \texttt{uranium} data set from the \textsf{R} package \texttt{copula} \citep{hofert2015copula} and decided that it could not be modelled by a simplified vine copula. \cite{killiches2015model} developed a test for the simplifying assumption and applied it to the same data set obtaining similar results. In the test they used the \textsf{R} package \texttt{gamCopula} \citep{vatter2015gamCopula}, which provides an algorithm for the estimation of non-simplified vine copula models with the help of generalised additive models. A critical examination of the simplifying assumption was presented in \cite{spanhel2015simplified}, where the focus was on possible misspecifications of simplified vine copulas when the underlying true model is non-simplified. This paper contributes to this discussion by focusing on the visual implications of the simplifying assumption.\\

In the following sections we will use the parametric bivariate copula families implemented in \texttt{VineCopula} as building blocks. This group of copulas includes Gaussian ($\Nc$), Clayton ($\Cc$), Gumbel ($\Gc$), Frank ($\Fc$), Joe ($\Jc$), Clayton-Gumbel ($\BBa$), Joe-Frank ($\BBd$), Tawn type 1 ($\Tc_{(1)}$) and Tawn type 2 ($\Tc_{(2)}$) copulas as well as their survival versions and rotations by $90$ degrees and $270$ degrees (indicated by the superscripts $180$, $90$ or $270$, respectively). The densities of the survival and rotated versions of a bivariate copula density $c$ are given by $c^{90}(u_1,u_2)=c(1-u_2,u_1)$, $c^{180}(u_1,u_2)=c(1-u_1,1-u_2)$ and $c^{270}(u_1,u_2)=c(u_2,1-u_1)$. When we specify a pair-copula, we state both the family and the corresponding parameters. For example, a Gaussian copula with correlation $\rho=0.5$ is denoted by $\Nc(0.5)$ and $\Tc_{(2)}^{270}(-3,0.6)$ stands for a Tawn type 2 copula rotated by $270$ degrees with first parameter $-3$ and second parameter $0.6$.

The space of admissible parameters depends on the copula family. For example, whereas the parameter space of a Tawn type 1 copula is $(1,\infty)\times (0,1)$, that of a Frank copula is given by $\Rbb\setminus \left\lbrace 0\right\rbrace $. Since we still want to compare different copula families we transform the parameters to the same scale using Kendall's $\tau$ as a measure for the strength of dependence. See for example \cite[Ch.\ 5.1.1]{nelsen2006introduction} for a discussion of Kendall's $\tau$ in the context of copulas.

\section{Visualisation of simplified vine copulas}\label{sec:Vis_Simp}

The contour of a density $f\colon\Rbb^d\to[0,\infty)$ corresponding to a level $y \in (0,\infty)$ is the set $\left\lbrace \zb \in \Rbb^d \mid f(\zb)=y \right\rbrace $ of all points in $\Rbb^d$ that are assigned the same density value $y$. For bivariate densities, plots of contour lines are well-known; in three dimensions this concept can be extended to contour surfaces. In this section we present contour plots of various simplified three-dimensional vine copula densities, ranging from very simple models such as a Gaussian copula, to more complex scenarios. The main goal is to get a feeling for what simplified vine copulas look like in order to properly compare them to non-simplified vine copulas. As well as the three-dimensional contour surfaces plotted from three different angles we present the contour lines of the two-dimensional marginals $c_{12}$, $c_{23}$ and $c_{13}$. While $c_{12}$ and $c_{23}$ are explicitly specified in the vine copula construction, the margin $c_{13}$ has to be calculated by integrating $c_{123}$ with respect to $u_2$, either analytically (when possible) or numerically as noted in equation \eqref{eq:c13}. 

For all two- and three-dimensional contour plots we take the univariate marginals to have a standard normal distribution, i.e.\ we consider the random vector $\Zb=(Z_1,Z_2,Z_3)^\top$, where $Z_j=\Phi^{-1}(U_j)$, $j=1,2,3$, with $\Phi$ denoting the standard normal distribution function. This is done because on the uniform scale copula densities would be difficult to interpret and hardly comparable with each other. Further, in this way a Gaussian copula corresponds to a Gaussian distribution so that all examples can be seen in comparison to this well-known case.

For interested readers we refer to our web-application (\texttt{\url{https://vinecopula.shinyapps.io/Vine3DPlot}}), where three-dimensional vine copula objects can be generated and rotated at one's convenience. This might be interesting if you wish to examine contour plots of our scenarios from other angles or if you are interested in visualising your own examples.

In \autoref{sec:Vis_Simp} we will consider the simplified vine copula specifications from \autoref{tab:scenarios} (Scenarios 1 to 4). Later, in \autoref{sec:Vis_Non_Simp}, we will also examine non-simplified vine copulas specifications and their simplified vine copula approximations. These are described in Scenarios 5 to 8 in \autoref{tab:scenarios}. For each scenario the three pair-copulas are specified by their families and parameter(s). Further we state the corresponding Kendall's $\tau$ value in order to facilitate comparability.

\begin{table}[htbp]
	\caption{Vine copula specifications considered in \autoref{sec:Vis_Simp} (simplified, Scenarios 1 to 4) and \autoref{sec:Vis_Non_Simp} (non-simplified, Scenarios 5 to 8). In Scenario 7, $\text{AMH}$ stands for the Ali-Mikhail-Haq copula. For a definition we refer to \cite{kumar2010probability}.}
	{\centering\small
		
		\begin{tabular}{cc|cccc|cccc}
			\rule[-2ex]{0pt}{5.5ex} &  & \multicolumn{4}{c|}{copula $c_{12}$} & \multicolumn{4}{c}{copula $c_{23}$} \\
			\cline{3-10}
			\rule[-2ex]{0pt}{5.5ex}  scenario & page  &  family & $\theta^{(1)}_{12}$ & $\theta^{(2)}_{12}$ & $\tau_{12}$ & family & $\theta^{(1)}_{23}$ & $\theta^{(2)}_{23}$ & $\tau_{23}$ \\
			\hline \rule[-2ex]{0pt}{5.5ex} 1 & \pageref{par:S_Gauss} & $\Nc$ & $0.6$ & $-$ & $0.41$ & $\Nc$ & $0.7$ & $-$ & $0.49$\\
			\hline \rule[-2ex]{0pt}{5.5ex} 2 & \pageref{par:S_Clayton} & $\Cc$ & $2$ & $-$ & $0.50$ & $\Cc$ & $2$ & $-$ & $0.50$ \\
			\hline \rule[-2ex]{0pt}{5.5ex} 3 & \pageref{par:S_mixed1} & $\Fc$ & $7$ & $-$ & $0.56$ & $\Gc$ & $2$ & $-$ & $0.50$\\
			\hline \rule[-2ex]{0pt}{5.5ex} 4 & \pageref{par:S_mixed2} & $\Tc_{(1)}$ & $3$ & $0.3$ & $0.25$& $\Jc^{270}$ & $-2$ & $-$ & $-0.36$\\
			\hline \rule[-2ex]{0pt}{5.5ex} 5 & \pageref{par:NS_Gauss} & $\Nc$ & $0$ & $-$ & $0$ & $\Nc$ & $0$ & $-$ & $0$  \\
			\hline \rule[-2ex]{0pt}{5.5ex} 6 & \pageref{par:NS_Clayton} & $\Cc$ & $-2$ & $-$  & $-0.50$ & $\Cc$ & $2$ & $-$  & $0.50$  \\
			\hline \rule[-2ex]{0pt}{5.5ex} 7 & \pageref{par:NS_Frank} & $\Fc$ & $8$ & $-$  & $0.60$ & $\Fc$ & $8$ & $-$  & $0.60$ \\
			\hline \rule[-2ex]{0pt}{5.5ex} 8 & \pageref{par:NS_mixed_high} & $\BBd$ & $6$ & $0.95$ & $0.69$ & $\Gc^{270}$ & $-3.5$ & $-$ & $-0.71$ \\
		\end{tabular}
		\newline
		\vspace*{5mm}
		\newline
		\begin{tabular}{cc|cccc}
			\rule[-2ex]{0pt}{5.5ex} &  & \multicolumn{4}{c}{copula $c_{13;2}$}  \\
			\cline{3-6}
			\rule[-2ex]{0pt}{5.5ex}  scenario & page  &  family & $\theta^{(1)}_{13;2}(u_2)$ & $\theta^{(2)}_{13;2}(u_2)$ & $\tau_{13;2}(u_2)$\\
			\hline \rule[-2ex]{0pt}{5.5ex} 1 & \pageref{par:S_Gauss} &  $\Nc$ & $0.5$ & $-$ & $0.33$\\
			\hline \rule[-2ex]{0pt}{5.5ex} 2 & \pageref{par:S_Clayton} & $\Cc$ & $0.67$ & $-$ & $0.25$ \\
			\hline \rule[-2ex]{0pt}{5.5ex} 3 & \pageref{par:S_mixed1} & $\Nc$ & $-0.7$ & $-$ & $-0.49$\\
			\hline \rule[-2ex]{0pt}{5.5ex} 4 & \pageref{par:S_mixed2} & $\BBa$ & $2$ & $1.5$ & $0.67$\\
			\hline \rule[-2ex]{0pt}{5.5ex} 5 & \pageref{par:NS_Gauss} &$\Nc$ & $0.9\sin(2\pi u_2)$ & $-$ & $\tau_{13;2}^{(5)}(u_2)$\\
			\hline \rule[-2ex]{0pt}{5.5ex} 6 & \pageref{par:NS_Clayton} & $\Cc$ & $9(-(u_2-0.5)^2+0.25)$ & $-$ & $\tau_{13;2}^{(6)}(u_2)$\\
			\hline \rule[-2ex]{0pt}{5.5ex} 7 & \pageref{par:NS_Frank} & $\text{AMH}$
			& $1-\exp(-8u_2)$ & $-$ & $\tau_{13;2}^{(7)}(u_2)$\\
			\hline \rule[-2ex]{0pt}{5.5ex} 8 & \pageref{par:NS_mixed_high} &  $\Tc_{(2)}/\Tc_{(2)}^{90}$ & $\sgn(u_2-0.5)(4-3\cos(8\pi u_2))$ & $0.1+0.8u_2$ & $\tau_{13;2}^{(9)}(u_2)$\\
		\end{tabular}
		\label{tab:scenarios}
	}
\end{table}

\subsection{Gaussian copula}\label{par:S_Gauss}
The first scenario we consider concerns a Gaussian copula. Among others, \cite{stoeber2013simplified} showed that every Gaussian copula can be represented as a simplified \textit{Gaussian vine copula} (i.e.\ all pair-copulas are Gaussian) and vice versa. 
We specify the pair-copulas of the vine as follows: $c_{12}$ is a bivariate Gaussian copula with parameter $\rho_{12}=0.6$ (i.e. $\tau_{12}=0.41$),  $c_{23}$ is a Gaussian copula with $\rho_{23}=0.7$ ($\tau_{23}=0.49$) and $c_{13;2}$ is a Gaussian copula with $\rho_{13;2}=0.5$ ($\tau_{13;2}=0.33$). 
This specification, which can be found in \autoref{tab:scenarios} (Scenario 1), directly implies that $c_{13}$ is a Gaussian copula with $\rho_{13}=0.71$ ($\tau_{13}=0.50$), see for example \cite{kurowicka2006uncertainty}, p.\ 69. The resulting elliptical-shaped contours displayed from three viewpoints in the top row of \autoref{fig:Sc1} are the natural extension of the well-known ellipsoid-shaped contour plots of bivariate normal distributions. We chose the contour levels for the plots such that the four contour surfaces are representative of the entire density. For the remainder of this paper these levels are fixed with values $0.015$, $0.035$, $0.075$ and $0.11$ (from outer to inner surface). The contour plots of the two-dimensional margins in the bottom row of \autoref{fig:Sc1} are those of bivariate normal distributions. We see that the contour plots of the bivariate margins already give a good impression of what the three-dimensional object looks like. It turns out that this property can be observed for all simplified vine copulas we will consider.

\begin{figure}[h!]
	\centering
	\includegraphics[width=1\linewidth]{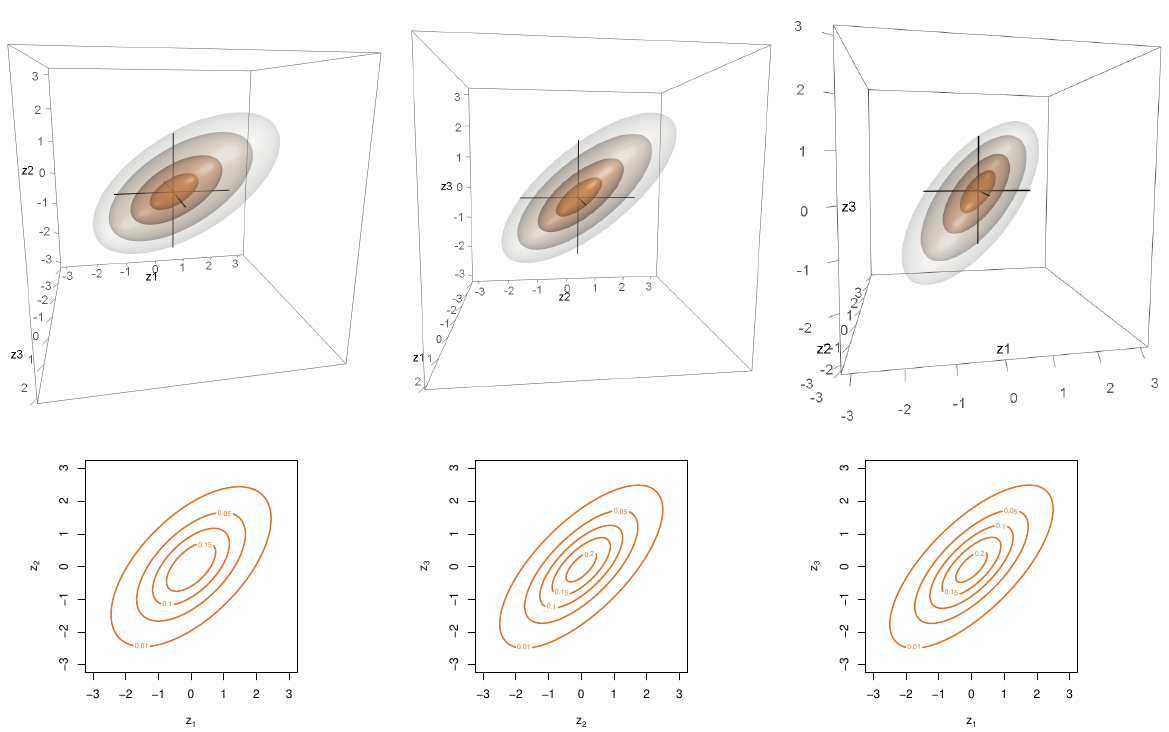}
	\caption{Top: Contours of the three-dimensional vine copula density specified by $c_{12}$: $\Nc(0.6)$, $c_{23}$: $\Nc(0.7)$, $c_{13;2}$: $\Nc(0.5)$. Bottom: Contours of the corresponding bivariate marginal densities.}
	\label{fig:Sc1}
\end{figure}

\subsection{Clayton copula}\label{par:S_Clayton}
A well-known representative of the class of Archimedean copulas is the Clayton copula. It is a one-parametric family with lower tail dependence. The Clayton copula is the copula underlying the multivariate Pareto distribution and is the only Archimedean copula that can be represented as a simplified vine copula as proved in \cite{stoeber2013simplified}, Theorem.\ 3.1. It is easy to see that the bivariate margins of a three-dimensional Clayton copula with parameter $\theta$ are bivariate Clayton copulas with parameter $\theta$, see for example \cite{kraus2015d}, Appendix B. There it was also shown that the copula of the conditioned variables (in our decomposition $c_{13;2}$) again is a Clayton copula, in this case with parameter $\theta/(\theta+1).$ Hence, in order to obtain a three-dimensional Clayton copula with parameter $\theta=2$ we specify a vine copula as described in Scenario 2 of \autoref{tab:scenarios}. The top row of \autoref{fig:Sc2} displays the contours of the resulting copula, the strong lower tail dependence is clearly visible. As already stated, the (unconditional) bivariate margin $c_{13}$ is also a Clayton copula with parameter $2$ and therefore all contour plots of the margins in the bottom row of \autoref{fig:Sc2} are identical. Again we observe that the shape of the contours of the three-dimensional density is anticipated quite well already by the two-dimensional marginal plots.

\begin{figure}[h!]
	\centering
	\includegraphics[width=1\linewidth]{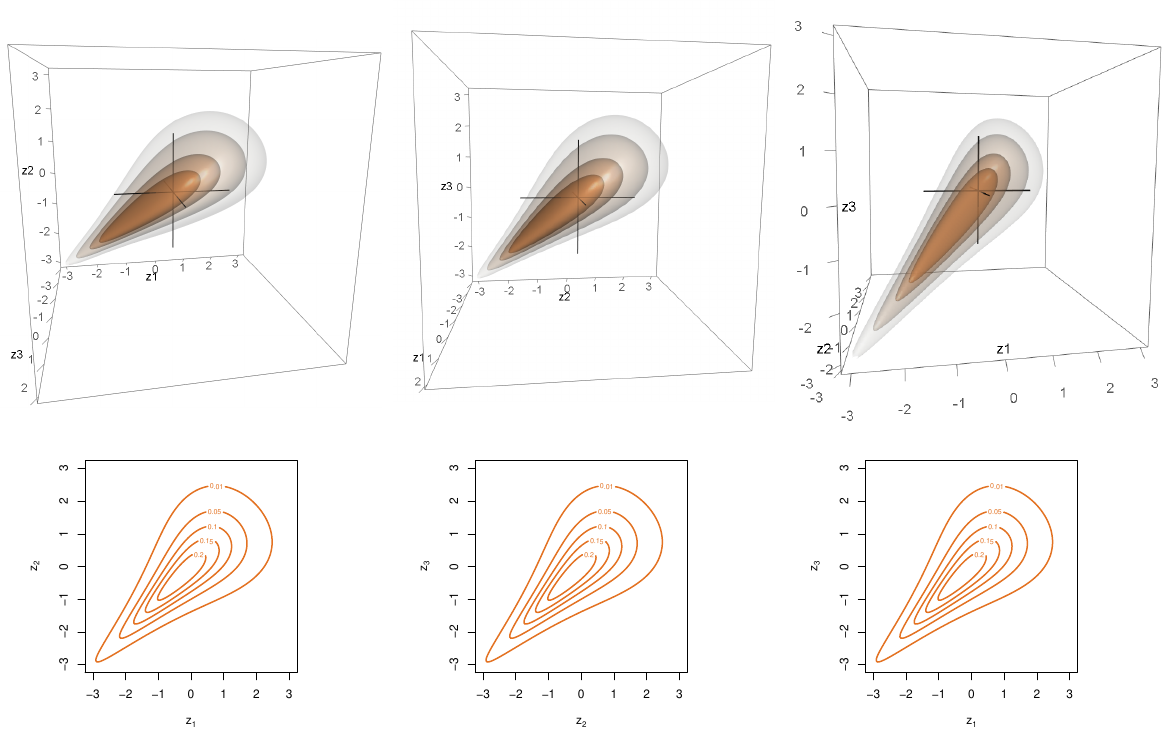}
	\caption{Top: Contours of the three-dimensional vine copula density specified by $c_{12}$: $\Cc(2)$, $c_{23}$: $\Cc(2)$, $c_{13;2}$: $\Cc(0.67)$. Bottom: Contours of the corresponding bivariate marginal densities.}
	\label{fig:Sc2}
\end{figure}

\subsection{Mixed simplified vine copula 1}\label{par:S_mixed1}
Up to now we have only considered vine copulas where all pair-copulas belong to the same family of parametric copulas. Of course, one of the main advantages of vine copulas is that one can specify each pair to be from a different copula family with its own parameter(s). The resulting model class is very flexible and able to describe many different kinds of dependencies. As an example for this, we present Scenario 3 (\autoref{tab:scenarios}): $c_{12}$ is a bivariate Frank copula with parameter $\theta_{12}=7$ (i.e. $\tau_{12}=0.56$), $c_{23}$ is a Gumbel copula with $\theta_{23}=2$ ($\tau_{23}=0.5$) and $c_{13;2}$ is a Gaussian copula with $\rho_{13;2}=-0.7$ ($\tau_{13;2}=-0.49$). 
In the resulting contour plots of \autoref{fig:Sc3} (top row) one can clearly see the shapes of the Frank and the Gumbel copula in the left and the middle plot, respectively. Although the dependency of each pair-copula is fairly strong, we observe rather weak dependence for $c_{13}$. The negative conditional dependence seems to cancel out with the positive dependencies implied by $c_{12}$ and $c_{23}$ (compare \autoref{fig:Sc3}, bottom row).

\begin{figure}[h!]
	\centering
	\includegraphics[width=1\linewidth]{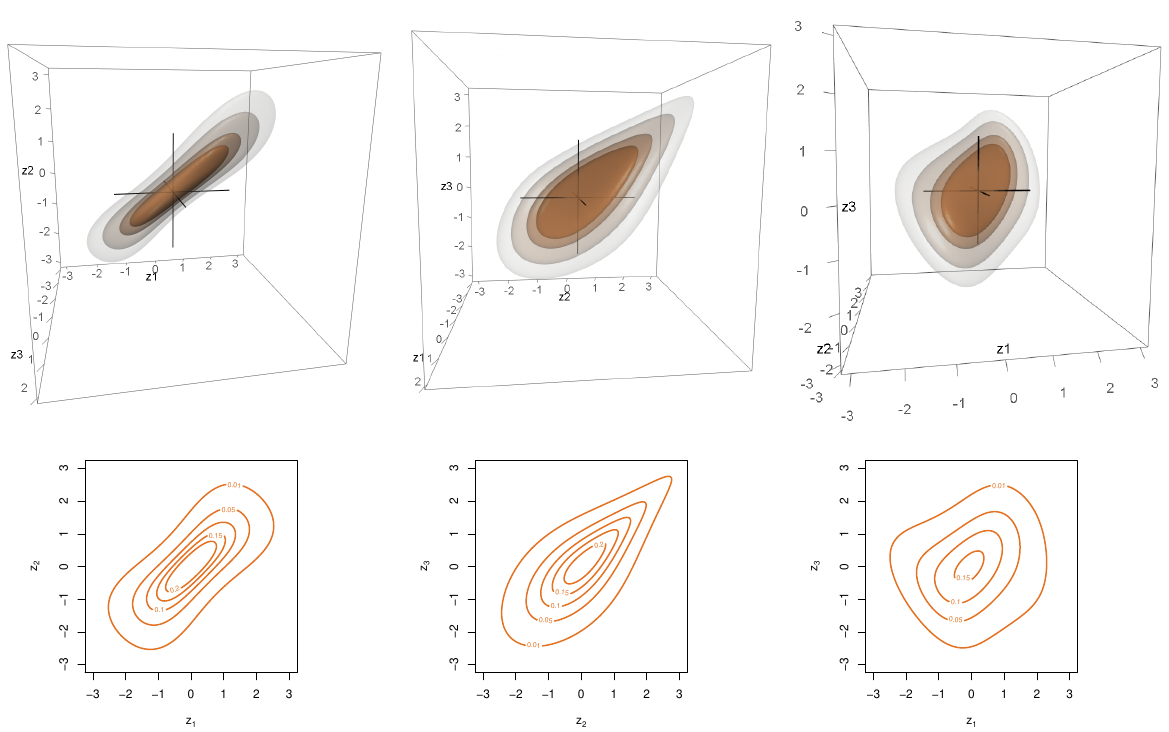}
	\caption{Top: Contours of the three-dimensional vine copula density specified by $c_{12}$: $\Fc(7)$, $c_{23}$: $\Gc(2)$, $c_{13;2}$: $\Nc(-0.7)$. Bottom: Contours of the corresponding bivariate marginal densities.}
	\label{fig:Sc3}
\end{figure}
\subsection{Mixed simplified vine copula 2}\label{par:S_mixed2}
\begin{figure}[h!]
	\centering
	\includegraphics[width=1\linewidth]{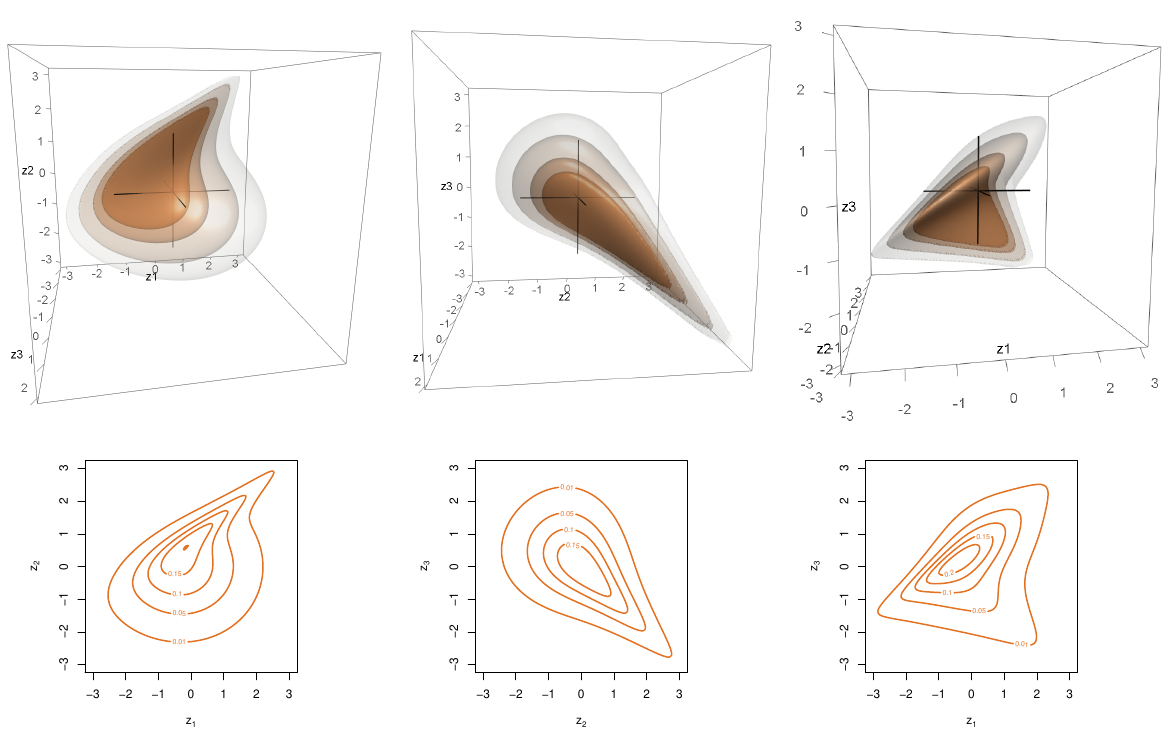}
	\caption{Top: Contours of the three-dimensional vine copula density specified by $c_{12}$: $\mathcal{T}_{(1)}(3,0.3)$, $c_{23}$: $\Jc^{270}(-2)$, $c_{13;2}$: $\BBa(2,1.5)$. Bottom: Contours of the corresponding bivariate marginal densities.}
	\label{fig:Sc4}
\end{figure}
We consider a second example of a mixed vine copula (Scenario 4 from \autoref{tab:scenarios}) with the following specifications: $c_{12}$ is a Tawn Type 1 copula with parameters $\thetab_{12}=(3,0.3)^\top$ (implying $\tau_{12}=0.25$), $c_{23}$ is a Joe copula rotated by $270$ degrees with $\theta_{23}=-2$ ($\tau_{23}=-0.36$) and $c_{13;2}$ is a $\BBa$ copula with $\thetab_{13;2}=(2,1.5)^\top$ ($\tau_{13;2}=0.67$). The shape of the resulting contours in the top row of \autoref{fig:Sc4} appears to be very non-standard. Especially the dependence between the first and third marginals is quite contorted. The dependence structure of the copula of the conditioned variables ($\BBa$) cannot be detected at all. Further the non-exchangeable nature of the Tawn copula is noticeable both in the three- and the two-dimensional contour plots (cf.\ \autoref{fig:Sc4}, bottom row).

Note that even for the rather bent examples in this section the shape of the bivariate marginal contour plots resembles what we see in the three-dimensional plots. Thus all considered simplified vine copulas share the property that knowledge of just the three bivariate margins already provides a fairly good idea of the shape of the contours of the three-dimensional copula density.

\section{Visualisation of non-simplified vine copulas}\label{sec:Vis_Non_Simp}

Having seen several examples of visualised simplified vine copulas we also aim to visually explore the meaning of the simplifying assumption. For this purpose we now present a series of contour plots of non-simplified vine copula densities and compare them to the ones of the corresponding simplified vine copula approximations. Similar to \cite{haff2010simplified} and \cite{stoeber2013simplified} we determine the simplified vine copula approximation of a non-simplified vine copula with pair-copulas $c^{\text{NS}}_{12}$, $c^{\text{NS}}_{23}$ and $c^{\text{NS}}_{13;2}$ by setting the unconditional pair-copulas $c^{\text{S}}_{12}$ and $c^{\text{S}}_{23}$ to the true ones ($c^{\text{NS}}_{12}$ and $c^{\text{NS}}_{23}$, respectively) and finding the pair-copula $c^{\text{S}}_{13;2}$ (independent of $u_2$) that minimises the Kullback-Leibler distance to the true conditional copula $c^{\text{NS}}_{13;2}$. Since in most scenarios considered in this chapter this minimisation is analytically infeasible, we estimate $c^{\text{S}}_{13;2}$ by generating a sample $\big(u^{(i)}_1,u^{(i)}_2,u^{(i)}_3\big)^\top_{i=1,\ldots,N}$ of size $N$ from the non-simplified model and fitting the likelihood maximising parametric bivariate copula $c^{\text{S}}_{13;2}$ to the copula data $\big(u^{(i)}_{1|2},u^{(i)}_{3|2}\big)^\top_{i=1,\ldots,N}$, where $u^{(i)}_{j|2}=C_{j|2}\big(u^{(i)}_j|u^{(i)}_2\big)$, $j=1,3$. Even though we found that the estimated parameters had converged up to the second digit for $N=10,\!000$ we used $N=100,\!000$ due to low computational effort.

In \autoref{sec:Vis_Non_Simp} we will consider the non-simplified vine copula specifications from \autoref{tab:scenarios} (Scenarios 5 to 8).

\subsection{Gaussian vine copula with sinusoidal conditional dependence function}\label{par:NS_Gauss}
\begin{figure}[h!]
	\centering
	\includegraphics[width=1\linewidth]{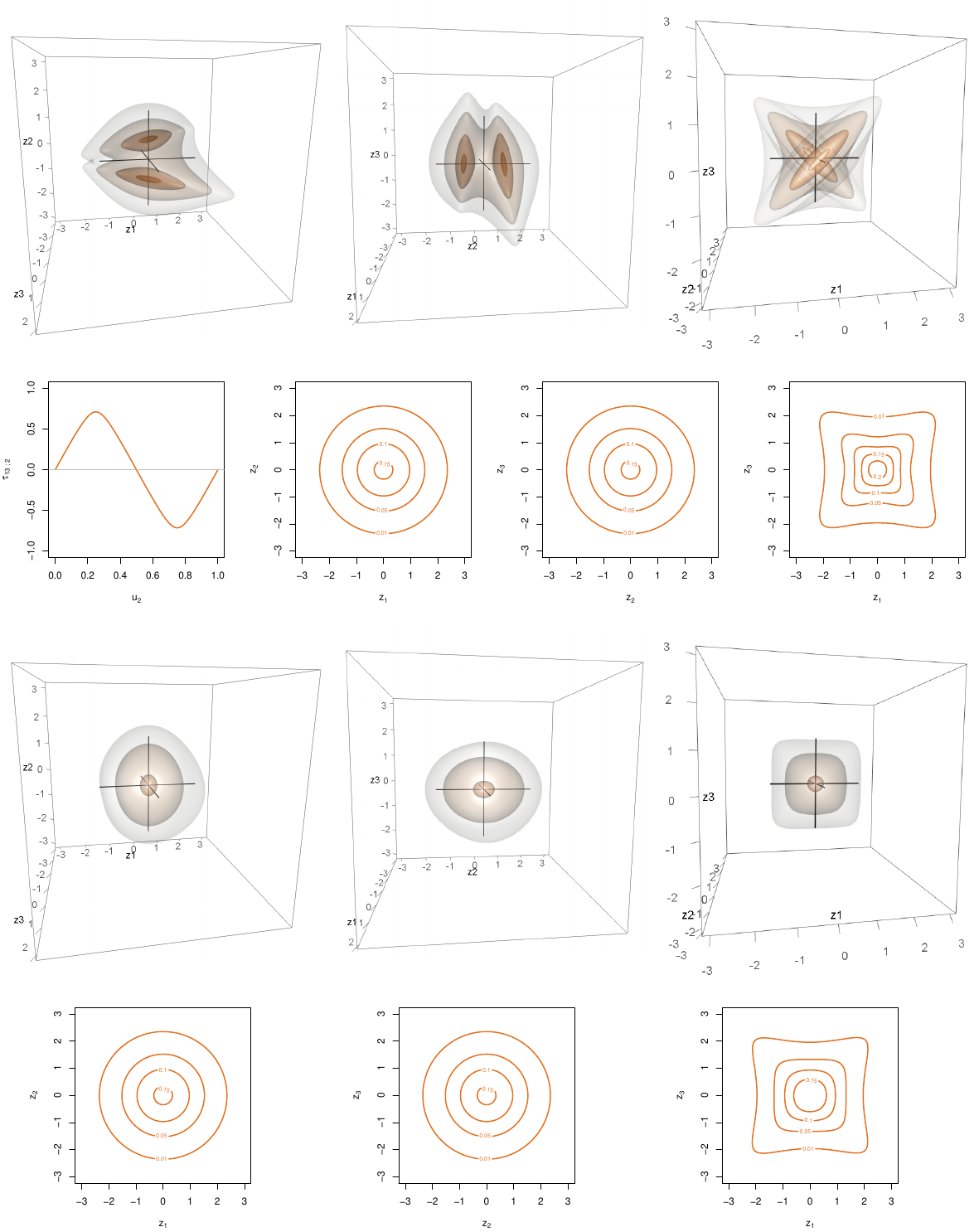}
	\caption{Top row: Contours of the three-dimensional non-simplified vine copula density specified by $c_{12}$: $\Nc(0)$, $c_{23}$: $\Nc(0)$, $c_{13;2}$: $\Nc(\rho_{13;2}(u_2))$ with $\rho_{13;2}(u_2)=0.9\sin(2\pi u_2)$.\\ Second row: $\tau_{13;2}$ depending on $u_2$ and contours of the bivariate margins corresponding to the specification of the top row.\\ Third row: Contours of the three-dimensional simplified vine copula approximation specified by $c_{12}$: $\Nc(0)$, $c_{23}$: $\Nc(0)$, $\hat c_{13;2}$: $\tcop(0.01,2.15)$.\\ Bottom row: Contours of the bivariate margins corresponding to the specification of the third row.}
	\label{fig:Sc5}
\end{figure}
In our first non-simplified example (Scenario 5 from \autoref{tab:scenarios}) we set the two unconditional copulas $c_{12}$ and $c_{23}$ to the independence copula in order to isolate the effect of the dependence of the conditional copula $c_{13;2}$ on $u_2$. We choose $c_{13;2}$ to be Gaussian with parameter function $\rho_{13;2}(u_2)=0.9\sin(2\pi u_2)$, i.e.\ one full period of a sine curve with amplitude $0.9$. Hence for values of $u_2$ ranging between $0$ and $0.5$ the dependence is positive with Kendall's $\tau$ between $0$ and $0.71$ and negative for $0.5<u_2<1$ (see also the left panel of the second row of \autoref{fig:Sc5}). The shift from positive to negative dependence is clearly visible in the contour plots shown in the top row of \autoref{fig:Sc5}. We also observe that the resulting contour surfaces for higher levels are no longer connected and the density is bimodal. Further, from the numerically integrated contour plot of $c_{13}$ in the right panel of the second row of \autoref{fig:Sc5} we conclude that marginally the strong positive and negative dependencies cancel each other out resulting in a bivariate marginal copula with almost no dependence, resembling a t copula with association of zero and low degrees of freedom.

In opposition to the simplified examples from \autoref{sec:Vis_Simp}, now the bivariate contour plots do no longer anticipate the three-dimensional object in a reasonable way. The sinusoidal structure of this copula cannot be guessed from the two-dimensional plots in the second row of \autoref{fig:Sc5}. In fact, had we used $\rho_{13;2}(u_2)=-0.9\sin(2\pi u_2)$, the copula density would have changed drastically (90 degree rotation along the z$_2$-axis) while the bivariate margins would have stayed exactly the same.

In contrast the corresponding simplified vine copula approximation, whose contours are displayed in the third row of \autoref{fig:Sc5}, resembles the smooth extension of the bivariate margins (bottom row of \autoref{fig:Sc5}) to three dimensions. This unimodal simplified vine copula, whose conditional copula $\hat c_{13;2}$ is indeed a t copula with almost no dependence ($\hat\rho_{13;2}=0.01$) and low degrees of freedom ($\hat\nu_{13;2}=2.15$), is not able to reproduce the twisted shape of the non-simplified vine copula at all. Also the implied bivariate margin of the first and third variable in the right panel of the bottom row of \autoref{fig:Sc5} is almost identical to the one of the non-simplified copula in the second row.

\subsection{Clayton vine copula with quadratic conditional dependence function}\label{par:NS_Clayton}
\begin{figure}[h!]
	\centering
	\includegraphics[width=1\linewidth]{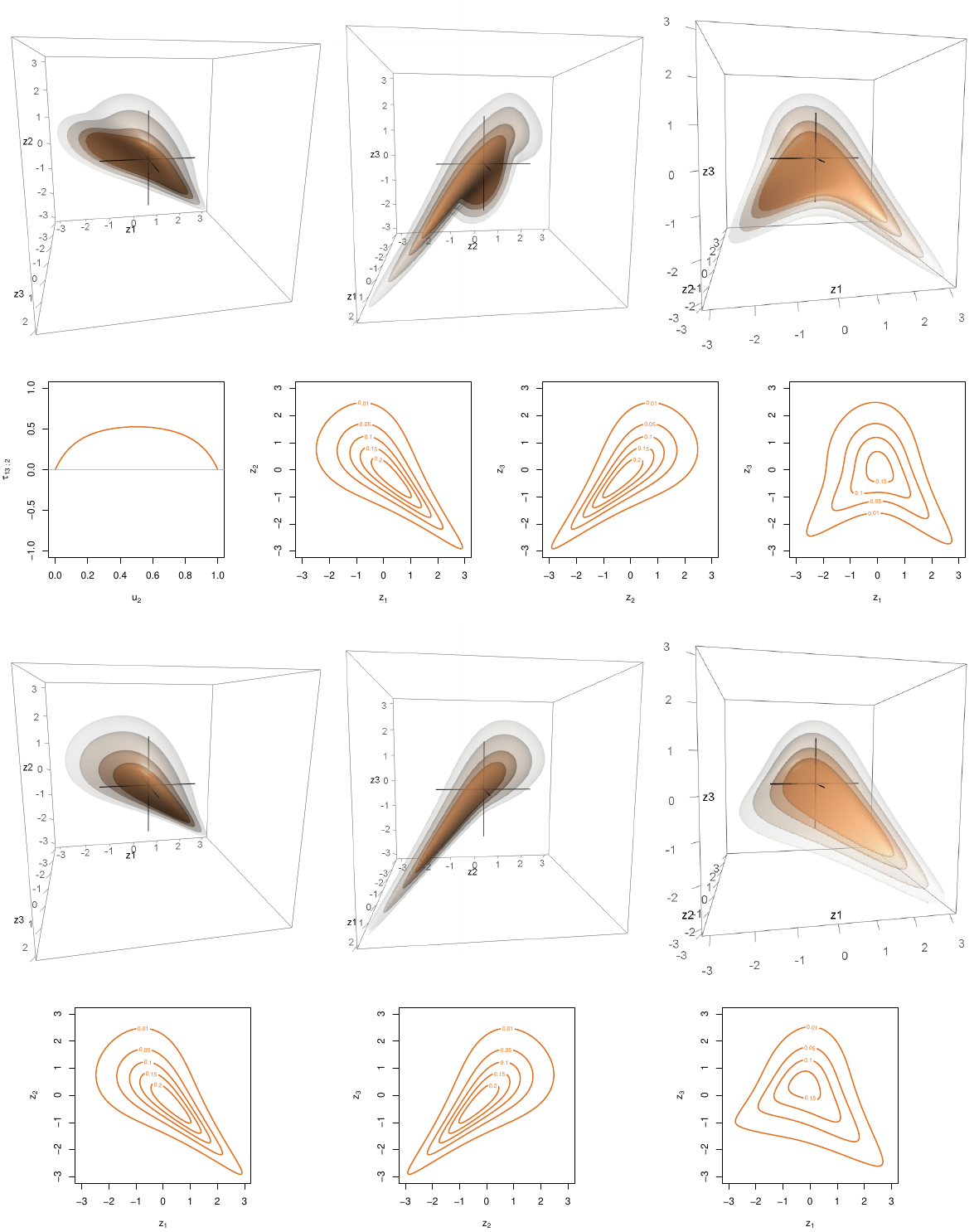}
	\caption{Top row: Contours of the three-dimensional non-simplified vine copula density specified by $c_{12}$: $\Cc^{90}(-2)$, $c_{23}$: $\Cc(2)$, $c_{13;2}$: $\Cc(\theta_{13;2}(u_2))$ with $\theta_{13;2}(u_2)=9(-(u_2-0.5)^2+0.25)$.\\ Second row: $\tau_{13;2}$ depending on $u_2$ and contours of the bivariate margins corresponding to the specification of the top row.\\ Third row: Contours of the three-dimensional simplified vine copula approximation specified by $c_{12}$: $\Cc^{90}(-2)$, $c_{23}$: $\Cc(2)$, $\hat c_{13;2}$: $\BBb^{180}(1.75,1.16)$.\\ Bottom row: Contours of the bivariate margins corresponding to the specification of the third row.}
	\label{fig:Sc6}
\end{figure}
Next we consider a non-simplified Clayton vine copula, i.e.\ all pair-copulas are bivariate Clayton copulas where the parameters of the unconditional copulas may differ in contrast to the three-dimensional Clayton copula (cf.\ Scenario 2) for which the parameters of $c_{12}$ and $c_{23}$ have to coincide. In this Scenario 6 we set the dependencies of the unconditional pair-copulas as $\theta_{12}=-2$ ($\tau_{12}=-0.5$) and $\theta_{23}=2$ ($\tau_{23}=0.5$) and specify the parameter function as a downwardly open parabola taking only non-negative values: $\theta_{13;2}(u_2)=9(-(u_2-0.5)^2+0.25)$. The corresponding $\tau_{13;2}$ values range between $0$ and $0.53$ and take their maximum for $u_2=0.5$ (see the left panel of the second row in \autoref{fig:Sc6}). The contours of the resulting density shown in the top row of \autoref{fig:Sc6} bear some resemblance to those of the Clayton copula (cf.\ \autoref{fig:Sc2}) but are much more distorted. Especially the relationship between the first and third variables seems to change from positive to negative dependence for different values of the second variable. This implies that also the conditional copula of $U_1$ and $U_3$ given $U_2=u_2$ exhibits a change from positive to negative dependence, which is an obvious indicator that the vine copula is non-simplified. The contours of the bivariate margin $c_{13}$ in the right panel of the second row of \autoref{fig:Sc6} also have a bent shape which is far from any of the standard parametric copulas. The bivariate contour plots of $c_{12}$ and $c_{23}$ suggest a smooth shape of the contours of the three-dimensional density such that one would not expect them to look as distorted as they do in the left and middle plots of the top row of \autoref{fig:Sc6}. 

Again the simplified vine copula approximation, which uses a survival BB6 copula with parameter $\hat\thetab_{13;2}=(1.75,1.16)$ as an approximation of the conditional copula $c_{13;2}$ (implying a Kendall's $\tau$ of 0.39), exhibits exactly this smooth behavior implied by the bivariate margins (see the third row of \autoref{fig:Sc6}). We further observe in the right plots of the last two rows of \autoref{fig:Sc6} that the simplified vine copula approximation is not able to reproduce the altering dependence pattern between $U_1$ and $U_3$ due to its constant conditional dependence parameter.

\subsection{Three-dimensional Frank copula}\label{par:NS_Frank}
In contrast to the Clayton copula, which can be expressed as a simplified vine copula (cf.\ \autoref{par:S_Clayton}), we now turn our attention to an Archimedean copula without this property, the three-dimensional Frank copula. Its non-simplified vine decomposition can be found as Scenario 7 in \autoref{tab:scenarios} (with dependence parameter $\theta=8$): The bivariate margins are again Frank copulas with the same dependence parameter $\theta$ (with corresponding Kendall's $\tau$ values equal to 0.6). The copula of the conditioned variables $c_{13;2}$ is also Archimedean, belonging to the Ali-Mikhail-Haq (AMH) family with functional dependence parameter $\gamma_{13;2}(u_2)=1-\exp(-\theta u_2)$ \citep[see][]{kumar2010probability,spanhel2015simplified}. The corresponding $\tau$ values displayed in the left panel of the second row of \autoref{fig:Sc7} show that the simplifying assumption is not severely violated. The strength of dependence is almost constant with the exception of small $\tau$ values for $u_2<0.2$. The contours depicted in the top row of \autoref{fig:Sc7} exhibit the typical bone shape known from the two-dimensional contour plots of bivariate Frank copulas such as those shown in the second row of \autoref{fig:Sc7}.
In order to assess how severe of a restriction the simplifying assumption would impose for modelling data generated by a Frank copula, we also present in the last two rows of \autoref{fig:Sc7} the three- and two-dimensional contours of the simplified vine copula approximation of the Frank copula, respectively.
For the trivariate Frank copula it is possible to analytically calculate the bivariate copula $c_{13;2}(\,\cdot\,,\cdot\,)$ that minimises the Kullback-Leibler divergence to the conditional copula $c_{13;2}(\,\cdot\,,\cdot\,;u_2)$, for details see \cite{spanhel2015simplified}. The visual difference between the Frank copula and its simplified vine copula approximation seems almost negligible. Only from the angle where the dependence between the first and third variable is visible the two contour plots can be distinguished (see the right plots of the first and third row of \autoref{fig:Sc2}). In the lower tail, where values of $z_2$ are small, the contours of the simplified vine copula approximation exhibit a higher dependence than the ones of the Frank copula, whose contours are less drawn into the corner implying less dependence. This is in line with what we would expect since the dependence function of $c_{13;2}$ of the non-simplified vine copula is decreasing for $u_2$ going to $0$, while the dependence of $c_{13;2}$ of the simplified vine copula approximation is constant at $\tau_{13;2}=0.28$.

\begin{figure}[h!]
	\centering
	\includegraphics[width=1\linewidth]{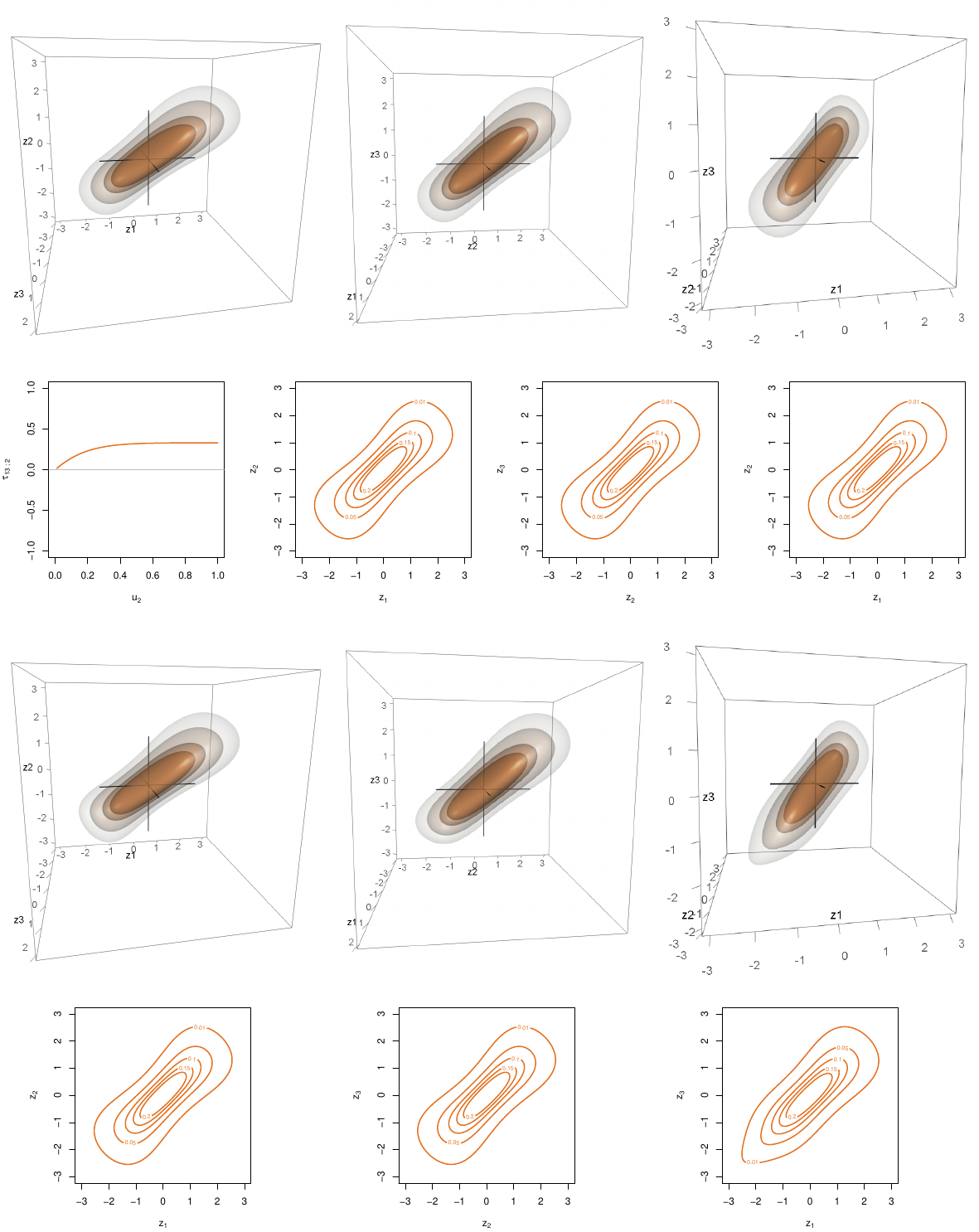}
	\caption{Top row: Contours of the three-dimensional non-simplified vine copula density specified by $c_{12}$: $\Fc(8)$, $c_{23}$: $\Fc(8)$, $c_{13;2}$: $\text{AMH}(\gamma_{13;2}(u_2))$ with $\gamma_{13;2}(u_2)=1-\exp(-8u_2)$.\\ Second row: $\tau_{13;2}$ depending on $u_2$ and contours of the bivariate margins corresponding to the specification of the top row.\\ Third row: Contours of the three-dimensional simplified vine copula approximation specified by $c_{12}$: $\Fc(8)$, $c_{23}$: $\Fc(8)$, $c_{13;2}$: see \cite{spanhel2015simplified}.\\ Bottom row: Contours of the bivariate margins corresponding to the specification of the third row.}
	\label{fig:Sc7}
\end{figure}

\subsection{Mixed non-simplified vine copula}\label{par:NS_mixed_high}
The last example we consider is Scenario 8 (see \autoref{tab:scenarios}). It is more extreme featuring pair-copulas with high dependence and more involved functions for the parameters of $c_{13;2}$. We specify $c_{12}$ as a $\BBd$ copula with parameters $\thetab_{12}=(6,0.95)^\top$ (implying $\tau_{12}=0.69$), $c_{23}$ as a Gumbel copula rotated by 270 degrees with $\theta_{23}=-3.5$ ($\tau_{23}=-0.71$) 
and $c_{13;2}$ as a Tawn Type 2 copula with both parameters depending on $u_2$ via the functions $\sgn(u_2-0.5)(4-3\cos(8\pi u_2))$ and $\theta_{13;2}^{(2)}(u_2)=0.1+0.8u_2$. The corresponding $\tau$ values ranging between $-0.39$ and $0.71$ are shown in the left panel of the second row of \autoref{fig:Sc9}. For the values of $u_2<0.5$ that imply negative dependence we use the 90 degree rotated version of the Tawn type 2 copula. \autoref{fig:Sc9} (top row) displays the contour plots of the resulting density. This is by far the most contorted density. The four peaks of the $\tau_{13;2}$ function are clearly visible as bumps in the three-dimensional contour plots. Of course, one can argue about how realistic it is to assume that real data follows such a distribution but it illustrates the variety of densities which can be modelled using non-simplified parametric vine copulas.

\begin{figure}[htbp!]
	\centering
	\includegraphics[width=1\linewidth]{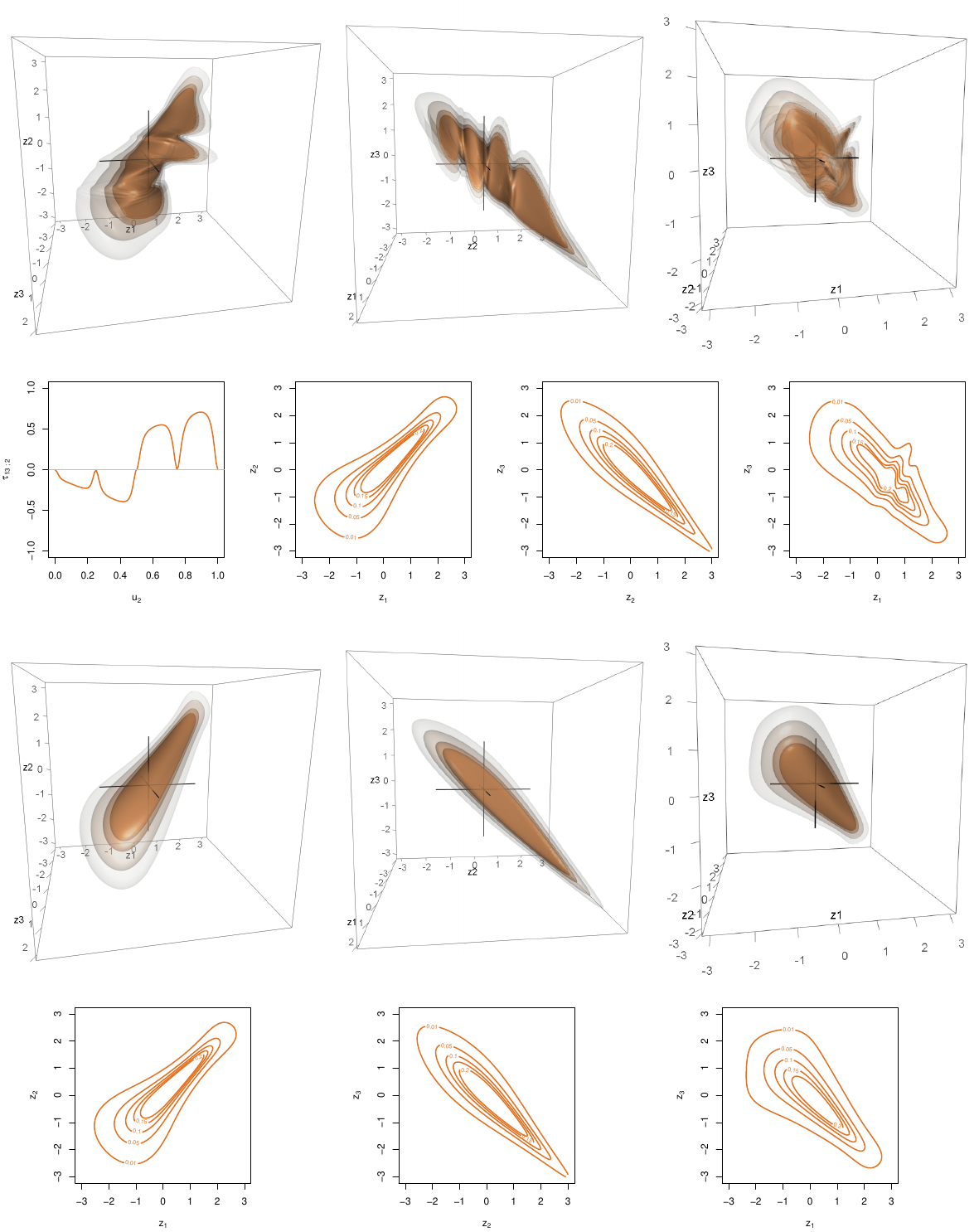}
	\caption{Top row: Contours of the three-dimensional non-simplified vine copula density specified by $c_{12}$: $\BBd(6,0.95)$, $c_{23}$: $\Gc^{270}(-3.5)$, $c_{13;2}$: $\mathcal{T}_{(2)}/\mathcal{T}_{(2)}^{90}(\theta_{13;2}^{(1)}(u_2),\theta_{13;2}^{(2)}(u_2))$ with $\theta_{13;2}^{(1)}(u_2)=\sgn(u_2-0.5)(4-3\cos(8\pi u_2))$ and $\theta_{13;2}^{(2)}(u_2)=0.1+0.8u_2$.\\ Second row: $\tau_{13;2}$ depending on $u_2$ and contours of the bivariate margins corresponding to the specification of the top row.\\ Third row: Contours of the three-dimensional simplified vine copula approximation specified by $c_{12}$: $\BBd(6,0.95)$, $c_{23}$: $\Gc^{270}(-3.5)$, $\hat c_{13;2}$: $\tcop(0.18,2.6)$.\\ Bottom row: Contours of the bivariate margins corresponding to the specification of the third row.}
	\label{fig:Sc9}
\end{figure}
For this scenario we can state that again the bivariate marginal contours do not really anticipate the complex shape of the corresponding three-dimensional object. The contour plots of $c_{12}$ and $c_{23}$ in the second row of \autoref{fig:Sc9} look perfectly smooth and regular and do not at all suggest the extremely twisted and contorted structure which can be seen in the top row of \autoref{fig:Sc9}. For the conditional copula of the corresponding simplified vine copula approximation a t copula with $\hat\rho_{13;2}=0.18$ and $\hat\nu_{13;2}=2.6$ is fitted, which corresponds to a Kendall's $\tau$ of 0.11. Comparing the resulting bivariate margins in the second and last row of \autoref{fig:Sc9} we see that apart from a little bump of $c_{13}$ of the non-simplified vine copula their general shapes are fairly similar. However the three-dimensional contour plot reveals that the simplified vine copula approximation (third row of \autoref{fig:Sc9}) is completely smooth with no twists and dents at all, such that it is not able to capture all aspects of the actual interdependencies.

\section{Application to simulated and real data}\label{sec:App}

In this section we want to investigate how the contour plots can help to decide whether a simplified or a non-simplified specification for given data is needed. For this purpose we at first consider simulated data, where we know the true underlying distribution, and afterwards apply the method to real data.

\subsection{Simulation study}\label{sec:A_simstudy}
For the simulated data example we specify the true non-simplified vine copula model as follows: We choose $c_{12}$ to be a Gumbel copula with parameter $\theta_{12}=1.5$ ($\tau_{12}=0.33$), $c_{23}$ as a t copula with $\rho_{23}=0$ and $2.5$ degrees of freedom ($\tau_{23}=0$) and $c_{13;2}$ as a Frank copula with parameter function $\theta_{13;2}(u_2)=3\arctan(10(u_2-0.5))$, implying negative dependence for $u_2<0.5$ and positive dependence for $u_2>0.5$ with absolute $\tau$ values smaller than $0.4$ (compare \autoref{fig:simstudy_2D}, top left panel). The rather low pairwise dependencies of this copula are clearly visible from the density's contour plots in the top row of \autoref{fig:simstudy}. However the surfaces look quite crumpled with lots of irregular bumps and deformations. Moreover it is eye-catching that in this example we only observe three contour surfaces. The inner surface is missing since the density only take values between $0$ and $0.101$ but the level of the inner surface is $0.11$. Further due to the low dependence we cannot detect any corner with extraordinarily high probability mass.

We generated a sample of size $N=3,\!000$ from this model and transformed the margins to be standard normal in order to make results comparable. For this transformed data sample, we performed a standard kernel density estimation with the function \texttt{kde} from the \textsf{R} package \texttt{ks} \citep{duong2016ks} using Gaussian kernels. Note that using this method we only get approximately standard normal margins. The contours of the resulting estimated densities, which are shown in the second row of \autoref{fig:simstudy}, are very close to those of the true underlying density in the top row. Only the innermost contour surface is smaller because the peaks of the density tend to get averaged out by kernel density estimation. The second row of \autoref{fig:simstudy_2D} displays the contours of the corresponding kernel density estimated bivariate margins, which are again close to the true ones in the first row of \autoref{fig:simstudy_2D}.

\begin{figure}[htbp]
	\centering
	\includegraphics[width=0.93\linewidth]{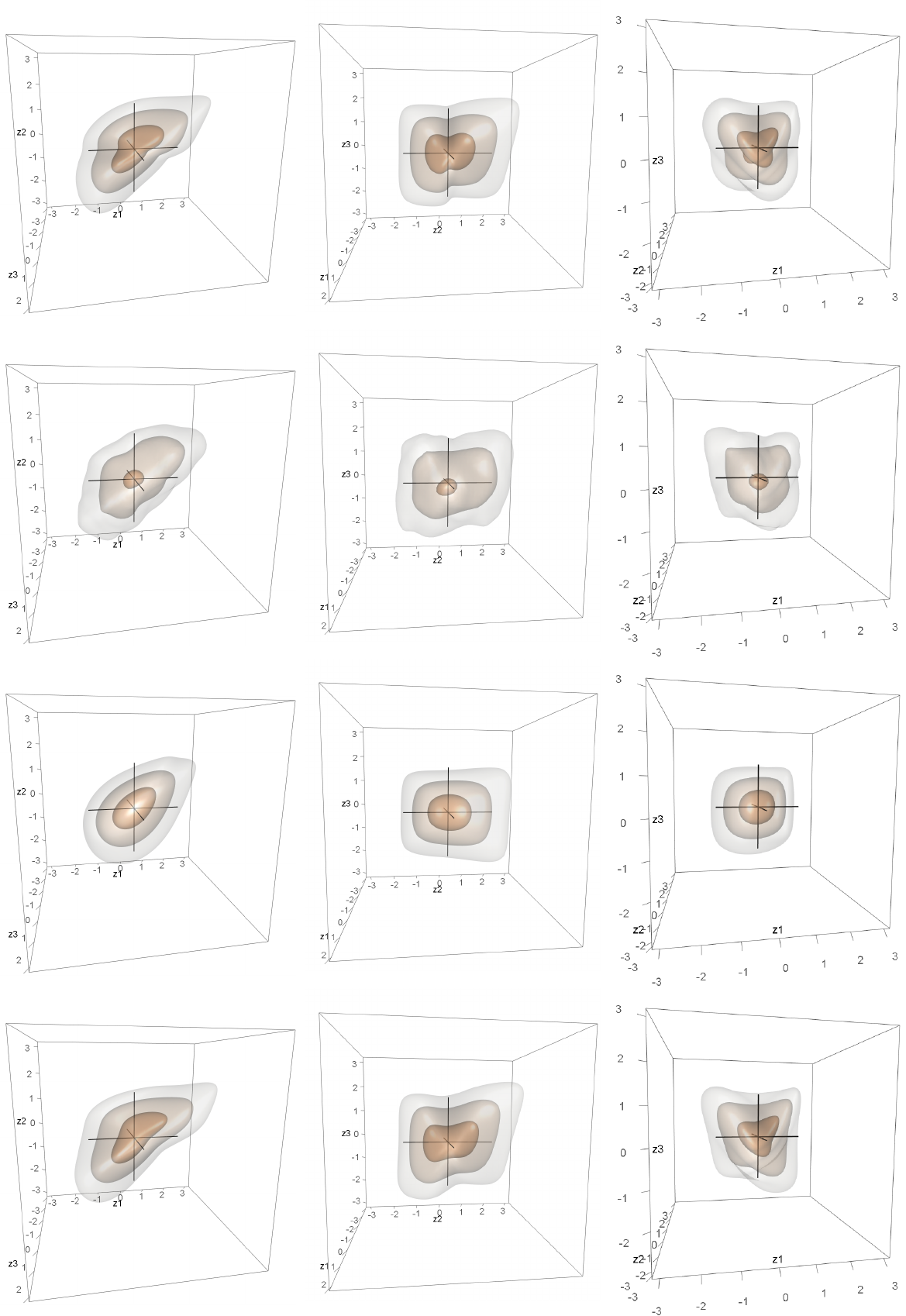}
	\caption{Top row: Contours of the true three-dimensional non-simplified vine copula density specified by $c_{12}$: $\Gc(1.5)$, $c_{23}$: $\tcop(0,2.5)$, $c_{13;2}$: $\Fc(\theta_{13;2}(u_2))$ with $\theta_{13;2}(u_2)=3\arctan(10(u_2-0.5))$.\\
		Second row: Contours of the density estimated via three-dimensional kernel density estimation.\\
		Third row: Contours of the fitted simplified vine copula density specified by $\hat c_{12}$: $\Gc(1.49)$, $\hat c_{23}$: $\tcop(0.04,2.36)$, $\hat c_{13;2}$:  $\tcop(0, 9.14)$.\\
		Bottom row: Contours of the fitted non-simplified vine copula density specified by $\hat c_{12}$: $\Gc(1.49)$, $\hat c_{23}$: $\tcop(0.04,2.36)$, $\hat c_{13;2}$: $\Nc(\hat{\rho}_{13;2}(u_2))$.}
	
	\label{fig:simstudy}
\end{figure}

\begin{figure}[h!]
	\centering
	\includegraphics[width=1\linewidth]{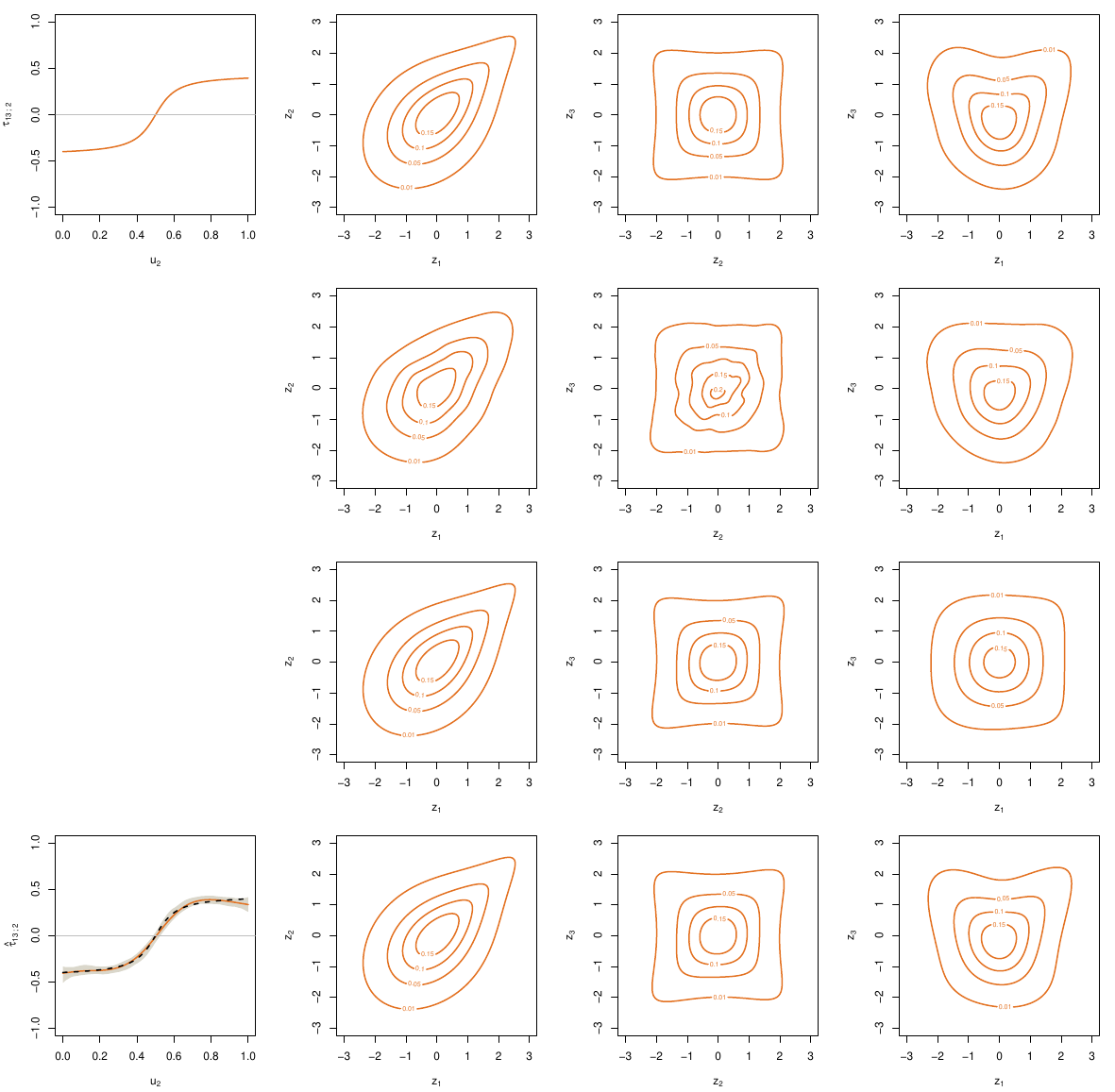}
	\caption{$\tau_{13;2}$ depending on $u_2$ (first column) and contour plots of the bivariate margins $c_{12}$ (second column), $c_{23}$ (third column) and $c_{13}$ (fourth column). Top row: true vine copula. Second row: kernel density estimation. Third row: fitted simplified vine copula. Bottom row: fitted non-simplified vine copula.}
	\label{fig:simstudy_2D}
\end{figure}

The idea is now to compare these contour plots to those of estimated simplified and non-simplified vine copula densities. We use \texttt{RVineStructureSelect} (from \texttt{VineCopula}) to fit a simplified vine copula and \texttt{gamVineStructureSelect} (from \texttt{gamCopula}) to fit a non-simplified vine copula. Both algorithms estimate the same unconditional copulas: $c_{12}$ is fitted as a Gumbel copula with parameter $\hat{\theta}_{12}=1.49$ and $c_{23}$ as a t copula with $\hat{\rho}_{23}=0.04$ and $\hat{\nu}_{23}=2.36$ degrees of freedom. In the simplified setting $c_{13;2}$ is estimated to be a t copula with $\hat{\rho}_{23}=-0.01$ and $\hat{\nu}_{23}=3.42$. The corresponding contours are shown in the third row of \autoref{fig:simstudy}. They seem to be an over-smoothed version of the kernel estimated density contours. While the general strengths of dependence are represented fairly well, the simplified vine copula approximation does not feature the bumps and dents of the kernel density estimated surfaces. A look at the contours of the bivariate margins in the third row of \autoref{fig:simstudy_2D} reveals that the densities of the explicitly modelled margins $c_{12}$ and $c_{23}$ are fitted very well. The true copula families are chosen and the estimated parameters are close to the true values. However the contours of the implicitly defined margin $c_{13}$ are far from the ones of the kernel density estimate. This is another indicator for an insufficient fit resulting from the underlying simplifying assumption, which is in this case too restrictive.

We now investigate whether these deficiencies can be remedied by fitting a non-simplified vine copula to the simulated data. The algorithm \texttt{gamVineStructureSelect} estimates the copula $c_{13;2}$ to be Gaussian with parameter function $\hat{\rho}_{13;2}(u_2)$ depending on $u_2$ via the functional form displayed in the bottom left panel of \autoref{fig:simstudy_2D} (in terms of $\hat{\tau}_{13;2}$), together with its bootstrapped $95\%$-confidence intervals (grey) and the true $\tau_{13;2}$ curve (dashed line). The $\tau$-values range between $-0.36$ and $0.38$ with negative values for $u_2<0.5$ such that the estimated function is quite close to the true underlying $\tau$-function. Even though the wrong copula family is chosen for $c_{13;2}$ (Gaussian instead of Frank) the contours of the resulting non-simplified vine copula in \autoref{fig:simstudy} (bottom row) are very similar to the kernel estimated ones and fit their shape considerably better than the estimated simplified vine copula. Also the contours of the bivariate margin $c_{13}$ in the bottom right panel of \autoref{fig:simstudy_2D} now provide a much better fit. Hence we can conclude that in this example we are able to visually detect the violation of the simplifying assumption of the true distribution.

\subsection{Real data application}\label{sec:RealDataApp}

In the following section we want to apply this method to a real data example. We investigate the well-known \texttt{uranium} data set, which can be found in the \textsf{R} package \texttt{copula}. This data set consists of 655 chemical analyses from water samples from a river near Grand Junction, Colorado (USA). It contains the log-concentration of seven chemicals, where we will focus on the three elements cobalt ($X_1$), titanium ($X_2$) and scandium ($X_3$) that have already been examined regarding the simplifying assumption in \cite{acar2012beyond} and \cite{killiches2015model}. In order to obtain copula data we first apply the probability integral transform to the data using the empirical marginal distribution functions, i.e.\ the observations $x_{ji}$, $j=1,2,3$, $i=1,\ldots,N$, are transformed via the rank transformation
\[
u_{ji}=\frac{1}{N+1}\sum_{k=1}^{N}1_{\left\lbrace x_{jk}\leq x_{ji}\right\rbrace },
\] 
where $1_{\left\lbrace \cdot \right\rbrace }$ is the indicator function. Then we transform the data to have standard normal margins in accordance to the previous examples.

We now want to take a look at the ``true'' model and perform a kernel density estimation. In the top rows of \autoref{fig:uranium} and \autoref{fig:uranium_2D} the results of the three- and two-dimensional kernel density estimations are displayed, respectively. The three variables seem to be positively dependent. A few bumps and dents are noticeable. Next we explore how well estimated simplified and non-simplified vine copulas fit the data.

\begin{figure}[h!]
	\centering
	\includegraphics[width=1\linewidth]{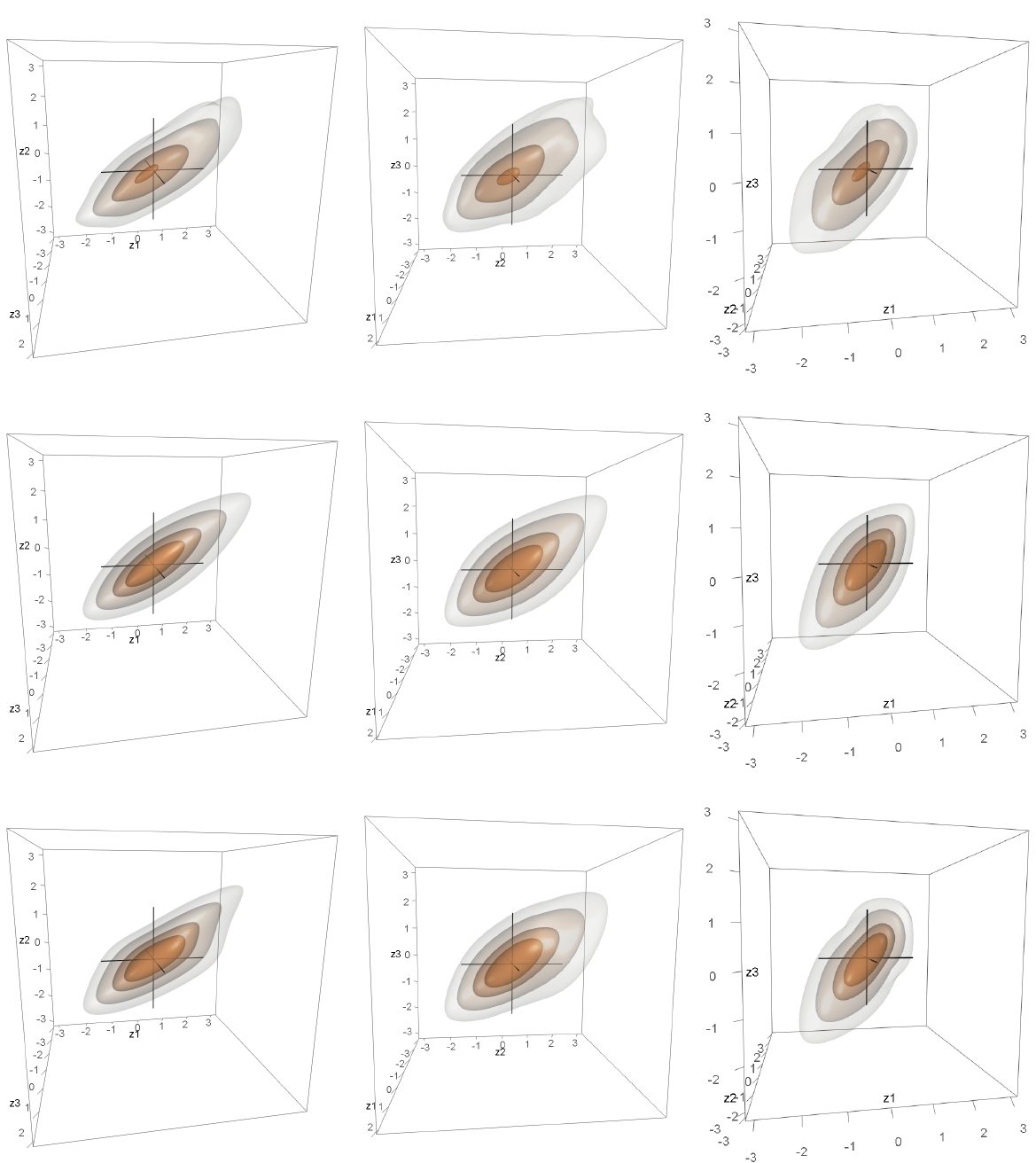}
	\caption{Top row: Contours of the density estimated via three-dimensional kernel density estimation.\\
		Middle row: Contours of the simplified vine copula specified by $\hat c_{12}$: $\tcop(0.53,8.03)$, $\hat c_{23}$:  $\tcop(0.43,5.93)$, $\hat c_{13;2}$:  $\tcop(0.08,5.65)$.\\
		Bottom row: Contours of the non-simplified vine copula specified by $\hat c_{12}$: $\tcop(0.53,8.03)$, $\hat c_{23}$:  $\tcop(0.43,5.93)$, $\hat c_{13;2}$: $\tcop(\hat{\rho}_{13;2}(u_2),6.69)$.}
	\label{fig:uranium}
\end{figure}

\begin{figure}[h!]
	\centering
	\includegraphics[width=1\linewidth]{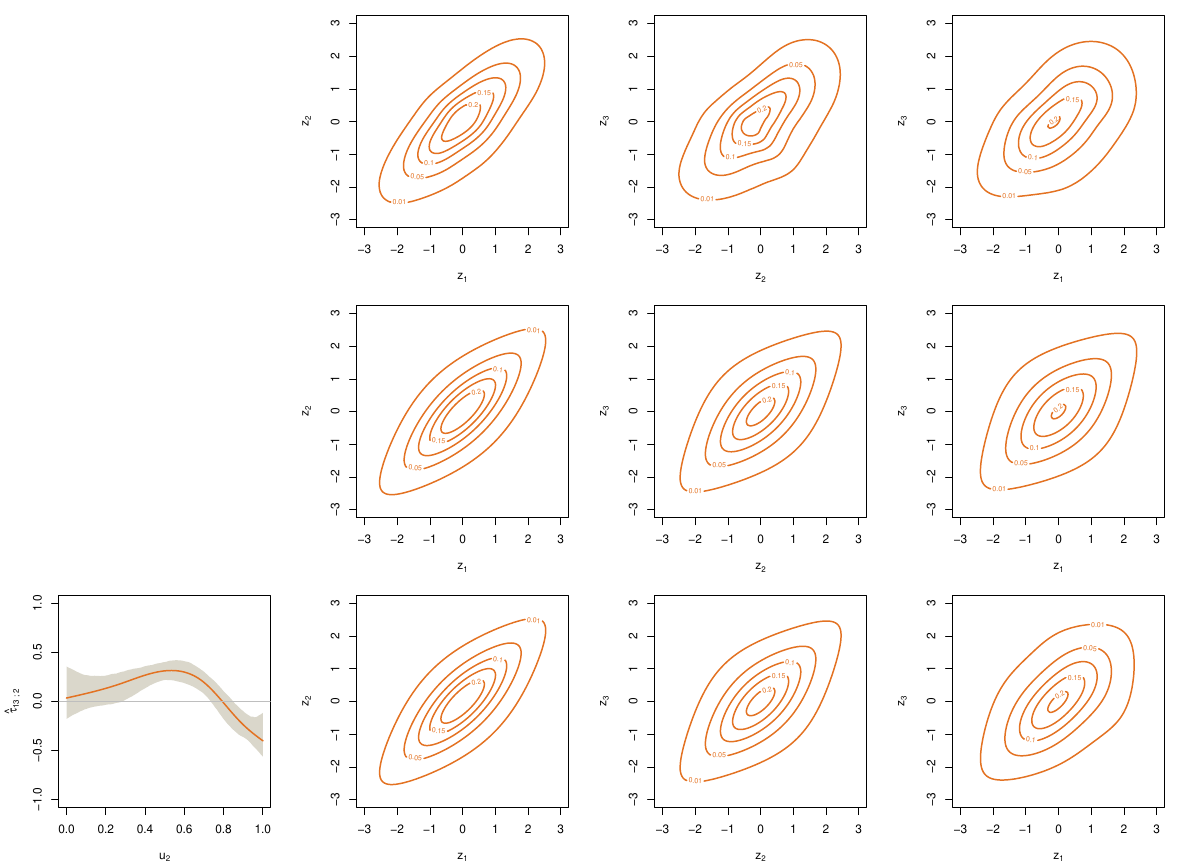}
	\caption{$\hat\tau_{13;2}$ depending on $u_2$ (bottom left) and contour plots of the bivariate margins $c_{12}$ (second column), $c_{23}$ (third column) and $c_{13}$ (fourth column). Top row: kernel density estimation. Middle row: estimated simplified vine copula. Bottom row: estimated non-simplified vine copula.}
	\label{fig:uranium_2D}
\end{figure}
\newpage 
Using \texttt{RVineStructureSelect} we obtain the following simplified vine copula: $c_{12}$ is a t copula with $\hat{\rho}_{12}=0.74$ and $\hat{\nu}_{12}=8.03$ ($\hat{\tau}_{12}=0.53$), $c_{23}$ is a t copula with $\hat{\rho}_{23}=0.63$ and $\hat{\nu}_{23}=5.93$ ($\hat{\tau}_{23}=0.43$) and $c_{13;2}$ is a t copula with $\hat{\rho}_{13;2}=0.08$ and $\hat{\nu}_{13;2}=5.65$ ($\hat{\tau}_{13;2}=0.05$). This t vine copula and its bivariate margins are depicted in \autoref{fig:uranium} (middle row) and \autoref{fig:uranium_2D} (middle row), respectively. Since all three degrees of freedom are of medium size we observe modest lower and upper tail dependence. Again these contours resemble a smoothed version of the slightly bumpy kernel density estimated contour surfaces resulting in a rather unsatisfying fit of the data.

For the non-simplified vine copula, the estimates of $c_{12}$ and $c_{23}$ are the same as for the simplified one. The third pair-copula $c_{13;2}$ is still a t copula but now with $\hat{\nu}_{13;2}=6.69$ degrees of freedom and an association parameter depending on $u_2$. We show the relationship between $u_2$ and $\hat{\tau}_{13;2}$ in the bottom left panel of \autoref{fig:uranium_2D} (again with its bootstrapped $95\%$-confidence intervals). One can see that we have small positive values of Kendall's $\tau$ for $u_2\leq 0.8$ and negative dependence for the remaining values of $u_2$. Although only the parameters of the copula $c_{13;2}$ are different compared to the simplified vine copula, the shapes of the contour surfaces display some interesting changes: Especially in the bottom left and right panel of \autoref{fig:uranium}, we see that the smooth diamond-shaped contours from \autoref{fig:uranium} (middle row) have developed several dents. While the contour plots of $c_{12}$ and $c_{23}$ are the same as before, the one of $c_{13}$ exhibits some differences since it is no longer diamond-shaped.

Comparing these contours to the ones from the top rows of \autoref{fig:uranium} and \autoref{fig:uranium_2D} we see that the non-simplified vine copula is able to capture the behaviour of the data quite well. The most noticeable bumps and dents are reproduced and the bivariate contours resemble the kernel density estimated ones. Thus we come to the same conclusions as \cite{acar2012beyond} and \cite{killiches2015model}, namely that the vine copula decomposition of this three-dimensional data set is of the non-simplified form.

\section{Conclusion}\label{sec:Conclusion}

In this paper we looked at the contour surfaces of several three-dimensional simplified and non-simplified vine copulas. The flexibility of simplified vine copulas in comparison to standard elliptical and Archimedean copulas was demonstrated. Using the 12 different one- and two-parametric bivariate pair-copula families currently implemented in \texttt{VineCopula} for the construction of a simplified vine copula, the shape of the resulting contour surfaces may deviate considerably from the well-known ellipsoid-shaped contours of a Gaussian distribution. Considering non-simplified vine copulas facilitates the modelling of even more irregular contour shapes exhibiting twists, bumps and altering dependence patterns. In our examples we have observed that contemplating three-dimensional contour surfaces gives more insight into the trivariate dependence structure than only looking at the two-dimensional marginal contour lines. While the consideration of the three bivariate marginal contour plots already gives a good impression of the shape of the three-dimensional object for simplified vine copulas, one might be surprised how twisted and contorted some non-simplified three-dimensional densities appear if one had only seen the smooth bivariate contour plots. In simulated and real data applications we have seen that non-simplified vine copulas are able to fit data with complex dependencies very well. However, we have observed that the estimated simplified vine copulas still capture the main features of the data providing a more smooth fit. Thus, for practical applications, especially in higher dimensions (when the number as well as the dimension of the parameter functions increase, causing numerical intractability) it might be preferable to use simplified vine copulas. Thereby overfitting might be avoided while the main properties of the data such as correlations and tail behaviour are still well represented.

\bibliographystyle{apalike}
\bibliography{visualization}

\end{document}